# Establishing an independent measurement traceability for $^{60}$Co Air Kerma

**N. Cornejo Díaz[1], C. García Mulas, P. Avilés Lucas**

*CIEMAT, Metrology Laboratory for ionizing Radiation, Ave. Complutense 40 (E01-P0-D13), 28040 Madrid, Spain*

*E-mail*: nestorarmando.cornejo@ciemat.es

ABSTRACT: An independent air kerma traceability chain for $^{60}$Co radiation protection levels has been successfully established at the CIEMAT Ionizing Radiation Metrology Laboratory (LMRI-CIEMAT), based on the reference provided by the $^{137}$Cs primary standard. To achieve this, a secondary standard ionization chamber with an appropriate energy response was characterized, and its beam-quality correction factor, $k_{\mathrm{Co,Cs}}$, was accurately determined via Monte Carlo simulations using the EGSnrc code. The method's accuracy was validated through a comparison with long-term historical calibration data and peer-confirmed through CIEMAT's successful participation in the EURAMET.RI(I)-S19 supplementary comparison, recently officially published in the BIPM KCDB. The excellent agreement achieved with international reference values demonstrates the high robustness and traceability independence of this newly implemented methodology.

KEYWORDS: Dosimetry concepts and apparatus; Detector modelling and simulations (Interaction of radiation with matter); Detector alignment and calibration methods (lasers, sources, particle-beams).

[1] Corresponding author.

## Contents

---

# 1 Introduction

The Reference Laboratory for Gamma Radiation at Protection Levels (RLGR-PL) of the LMRI-CIEMAT has established a primary measurement standard for air kerma in $^{137}$Cs [1]. Its degree of equivalence with the international standard has been established through the BIPM Key comparison BIPM.RI(I)-K5 [2]. The primary standard consists of two graphite-walled ionization chambers with nominal volumes below 10 cm$^3$, designed for high air kerma rates reaching several Gy/h.

To cover the wide range of air kerma rates produced by the RLGR-PL beams, several secondary standard ionization chambers with varying volumes are employed (table 1). The traceability chain is thus structured in a sequential hierarchy: from the Primary Reference Standard to the PTW 32005, and subsequently to the PTW 32002 and PTW 32003 chambers. Due to their intrinsic design, all the ionization chambers used ensure the charged-particle equilibrium required for the energies involved. This configuration ensures that all laboratory beams are calibrated with secondary standards of appropriate sensitivity and energy response. Subsequently, operational quantities for radiological protection are derived from air kerma by applying the conversion coefficients specified in the ISO 4037-3 standard [3], complemented by calculations [4]. It should be noted that the reference fields in the RLGR-PL strictly comply with the standard beams conditions required for the application of these conversion coefficients.

**Table 1**. Secondary standards and radiation qualities at RLGR-PL of the LMRI-CIEMAT

| Ionization chamber | Nominal volume / cm$^3$ | Beam qualities |
|---|---|---|
| PTW 32005 (S.N.: 00047) | 30 | S-Cs |
| PTW 32002 (S.N.: 00345) | 1000 | S-Cs and S-Co |
| PTW 32003 (S.N.: 00134) | 10000 | S-Cs and S-Co |

A technical challenge arises in the laboratory's $^{60}$Co beams, where the highest air kerma rates are on the order of a few mGy/h. At these protection levels, the small volume of the primary standard chambers results in insufficient signal-to-noise ratios, preventing their direct use for calibration. To bridge this gap, traceability to the $^{60}$Co radiation quality was established by means of a transfer chamber calibrated against the $^{137}$Cs primary reference. The calibration coefficient for the PTW 32002 chamber at the $^{60}$Co beam quality ($N_{K,\mathrm{Co}}$) is derived from its reference calibration in $^{137}$Cs ($N_{K,\mathrm{Cs}}$) according to equation (1):

$$N_{K,\mathrm{Co}} = N_{K,\mathrm{Cs}} \,.\, k_Q \tag{1}$$

where $k_Q$ (or $k_{\mathrm{Co,Cs}}$) represents the beam-quality correction factor. This factor was determined via Monte Carlo simulations and further validated through a comparison with historical calibration data and participation in a EURAMET supplementary comparison. The remarkably flat energy response of the PTW 32002 chamber in the energy range of interest [5] ensures the robustness of this approach. The present study details these investigations in the following sections.

# 2 Methods

## 2.1 Definition of the correction factor $k_{\mathrm{Co,Cs}}$

Adapting the conceptual framework of the TRS-398 protocol [6] (originally developed for radiotherapy beams) to radiation protection levels, the correction factor $k_{\mathrm{Co,Cs}}$ is defined as the ratio of the air kerma calibration coefficients for the PTW 32002 (S.N. 00345) ionization chamber at the $^{60}$Co and $^{137}$Cs radiation qualities, as shown in equation (2):

$$k_{\mathrm{Co,Cs}} = N_{K,\mathrm{Co}}/N_{K,\mathrm{Cs}} \tag{2}$$

where each quality-specific calibration coefficient, $N_{K,Q}$, is explicitly given by equation (3):

$$N_{K,Q} = K_{\mathrm{air},Q}/M_Q \tag{3}$$

Here, $K_{\mathrm{air},Q}$ represents the air kerma at the reference point on the beam axis in the absence of the chamber, and $M_Q$ is the chamber reading with its reference point positioned at the same location. In experimental determinations, $M_Q$ denotes the instrument response corrected for all relevant influence quantities (e.g., atmospheric pressure, ambient temperature, ion recombination and electrometer calibration). Such corrections, however, are not required in numerical simulations.

## 2.2 Determination of the factor $k_{\mathrm{Co,Cs}}$ via Monte Carlo simulation

The response of the ionization chamber in terms of electrical charge can be expressed as:

$$M_Q = \frac{D_Q \,.\, m_{\mathrm{air}}}{\left(W/e\right)_Q} \tag{4}$$

where $D_Q$ is the absorbed dose in the air of the chamber cavity for quality $Q$, $m_{\mathrm{air}}$ is the mass of air within the cavity, and $(W/e)_Q$ is the mean energy required to produce ion pair in air. Since $(W/e)_Q$ remains constant for photons and electrons within the energy range of interest, the subscript $Q$ my be omitted.

By substituting equation (4) into equation (3) for both radiation qualities, the following expressions are obtained:

$$N_{K,\mathrm{Co}} = \frac{K_{\mathrm{air,Co}} \left(W/e\right)}{D_{\mathrm{Co}} \,.\, m_{\mathrm{air}}} \tag{5}$$

$$N_{K,\mathrm{Cs}} = \frac{K_{\mathrm{air,Cs}} \left(W/e\right)}{D_{\mathrm{Cs}} \,.\, m_{\mathrm{air}}} \tag{6}$$

When calculating the correction factor $k_{\mathrm{Co,Cs}}$ using equation (2), both the mass of air ($m_{\mathrm{air}}$) and the *W*/*e* value cancel out. Consequently, the exact volume of the chamber cavity is not required for the calculation, and the uncertainty associated with *W*/*e* does not contribute to the final uncertainty of $k_{\mathrm{Co,Cs}}$.

### 2.2.1 Air kerma calculation

The air kerma for each beam quality was determined by simulating a parallel photon beam incident on an air sphere (radius of 2 cm and density of 1.20479 g/cm³) placed in vacuum. Simulations were performed using the CAVITY user code of the EGSnrc software package [7]. The air kerma was calculated as the absorbed dose in air under Fano conditions, achieved by regenerating incident photons after each interaction and discarding scattered photons.

To enhance computational efficiency, a photon-splitting factor of 100 was applied. The electron cutoff energy (ECUT), defined as the sum of the rest mass and kinetic energy, was set to 1.173 MeV for $^{137}$Cs and 1.844 MeV for $^{60}$Co. These values were specifically chosen to encompass the maximum initial energy of the electrons released by the incident photons, thereby suppressing electron transport and subsequent bremsstrahlung losses. Consequently, the calculations effectively accounted only for the initial kinetic energy of the electrons, aligning with the definition of air kerma. Additionally, atomic relaxation processes were enabled to include the energy of emitted Auger electrons. Detailed transport parameters used in these calculations are summarized in table 2.

**Table 2**. Transport parameters used during the calculations of air kerma

| Parameters and cross sections | Options |
|---|---|
| Photon cross sections | xcom |
| Compton cross sections | default |
| Global Pcut / MeV | 0.01 |
| Pair cross sections | BH |
| Pair angular sampling | Simple |
| Triplet production | Off |
| Bound Compton scattering | norej |
| Radiative Compton corrections | Off |
| Rayleigh scattering | On |
| Atomic relaxations | eadl |
| Photoelectron angular sampling | On |
| Photonuclear attenuation | Off |
| Photonuclear cross sections | default |
| Global Ecut / MeV | 1.173 ($^{137}$Cs) or 1.844 ($^{60}$Co) |
| Brems cross sections | BH |
| Brems angular sampling | KM |
| Spin effects | On |
| Electron Impact Ionization | Off |
| Global Smax | default |
| ESTEPE | 0.05 |
| Ximax | 0.5 |
| Boundary crossing algorithm | Exact |
| Skin depth for BCA | 3 |
| Electron-step algorithm | EGSnrc |
| Calculation type | Fano |
| Calculation geometry | Sphere_in_vacuum |

To evaluate the uncertainty associated with interaction cross-sections, calculations were repeated using both MCDF-XCOM (renormalized photoelectric cross-sections for XCOM) and MCDF-EPDL (evaluated photon data library) libraries of the EGSnrc package. To validate the simulation method, air kerma was also calculated analytically for the primary emission lines of $^{137}$Cs and $^{60}$Co using the following expression:

$$K_{\mathrm{air}} = \Phi\, E \frac{\left(\frac{\mu_{\mathrm{en}}}{\rho}\right)_{\mathrm{air}}}{(1-\bar{g})} \quad (7)$$

where $\Phi$ is the photon fluence at the reference point, $E$ is the photon energy, $(\mu_{\mathrm{en}}/\rho)_{\mathrm{air}}$ is the mass energy-absorption coefficient of air at energy $E$, and $\bar{g}$ is the average fraction of secondary electron kinetic energy lost by radiative processes in air.

Mass energy-absorption coefficients were obtained from ICRU Report 90 [8]. Calculations were performed using interpolated values from various cross-section datasets, including XCOM (with and without Scofield photoelectric cross-section renormalization) and the PENELOPE libraries. The $\bar{g}$ values (table 3) were computed using the "g" user code of the EGSnrc package. The uncertainties reported in table 3 represent the Type A standard uncertainty resulting from statistical fluctuations in the simulations.

**Table 3.** Values of $\bar{g}$ in air, for different photon energies

| Energía / keV | $\bar{g}$ | $u(\bar{g})$ (%)[(1)] |
|---|---|---|
| 661.17 | 0.00170 | 0.15 |
| 1173.2 | 0.00300 | 0.33 |
| 1332.5 | 0.00302 | 0.30 |

*[(1)] The standard uncertainties in $\bar{g}$ (k = 1) result in relative uncertainties of less than 0.01% in the $(1-\bar{g})$ term.*

### 2.2.2 Calculation of the absorbed dose in the air of the chamber cavity

Absorbed dose calculations within the chamber cavity were performed using Monte Carlo simulations with the CAVITY code of the EGSnrc system [7]. The mathematical model of the PTW 32002 chamber was developed based on specific information provided by the manufacturer, following the signing of a non-disclosure agreement [9]. Additionally, verification X-ray imaging with dimensional analysis was performed to complement the characterization of the chamber's internal geometry [10]. This information augments the generic data for the chamber model found in the user manual [5], which are summarized in table 4.

**Table 4**. Generic data for the PTW 32002 ionization chamber

| Chamber Feature | Description |
|---|---|
| Shape: | Spherical |
| Nominal volume: | 1000 $cm^3$ |
| Reference point: | Chamber center |
| Outer diameter: | 140 mm |
| Central electrode geometry: | Spherical |
| Central electrode diameter: | 50 mm |
| Central electrode material: | Graphite-coated polystyrene |
| Outer wall thickness: | 3 mm |
| Outer wall material: | POM $(CH_2O)_n$ with internal graphite coating |
| Wall surface density: | 453 $mg/cm^2$ |
| Energy dependence: | ≤ ± 1 % in the range from 200 keV to 1.33 MeV |

Figure 1 shows images of the PTW 32002 ionization chamber's mathematical model used during the simulations. Except for Global Ecut (0.521 MeV), Calculation type (Dose), and Calculation geometry (Chamber_in_vacuum), all transport parameters for the cavity absorbed dose calculations were identical to those specified in table 2.

To evaluate the simulation uncertainty arising from photon interaction cross sections, calculations were repeated using the mcdf-xcom ("renormalized photo-electric cross sections for XCOM") and mcdf-epdl ("renormalized photo-electric cross sections for evaluated photon data library") libraries of the EGSnrc package [7]. Furthermore, to assess the uncertainty arising from potential variations in the photon spectrum for each beam quality, air kerma and absorbed dose calculations were performed using three spectral sources. These sources included pure emission lines for $^{60}$Co and $^{137}$Cs [11], specific spectra from the laboratory irradiators [12, 13] and the $^{60}$Co spectrum reported by Mora et al. [14].

Additionally, to account for the uncertainty in absorbed dose arising from axial beam non-uniformity, resulting from finite source dimensions at different distances from the ionization chamber, simulations were performed for both parallel photon beams and point sources at different distances from the chamber's reference point. Furthermore, the uncertainty of the chamber model was evaluated by recalculating the factor $k_{\mathrm{Co,Cs}}$ focusing on the most critical geometric and material parameters. This conservative evaluation involved using a simplified model—excluding some details of the central electrode support—and applying maximum variations (± 10 %) to the outer wall mass thickness.

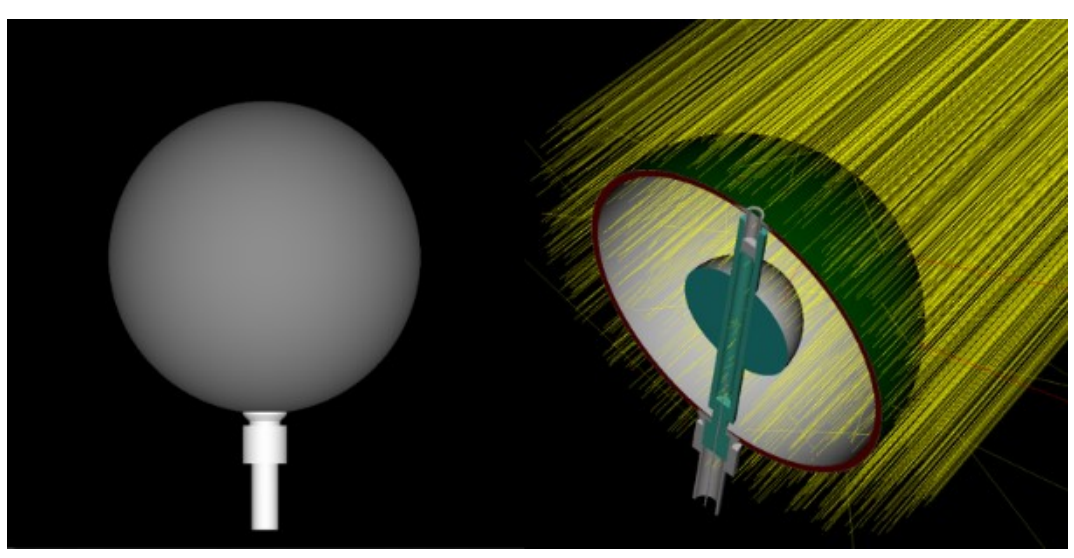

**Figure 1**. External view of the chamber's mathematical model (left) and cross-sectional view showing simulated radiation tracks (right). Images generated using the EGS_VIEW program [15].

## 3 Results and validation

### 3.1 The beam-quality correction factor $k_{\mathrm{Co,Cs}}$

The correction factor $k_{\mathrm{Co,Cs}}$ was calculated using equations (2), (5), and (6), based on the air kerma and the absorbed dose in the chamber air for the pure emission lines of $^{137}$Cs and $^{60}$Co, under parallel beam geometry. Table 5 summarizes the results, where $K_{\mathrm{air}}$, $D_Q$ and $M_Q$ are expressed per unit of photon fluence. The uncertainty values shown correspond to the relative standard uncertainty arising from statistical fluctuations during the simulations.

During absorbed dose calculations, the value for the air mass within the chamber was 1.4277 g, corresponding to a calculated collection volume of 1185 cm³ and an air density at reference conditions (Atmospheric pressure = 101.325 kPa and temperature = 20 °C) of 1.20479 × 10$^{-3}$ g cm$^{-3}$. The *W/e* value adopted follows the recommendation of ICRU Report 90 [8]. While these parameters cancel out in the calculation of the $k_{Co,Cs}$ factor, recommended values were employed to ensure that the individual calibration coefficients obtained for both beam qualities remained physically consistent. In this calculation, the uncertainties of *W/e* and the air mass in the chamber cavity are fully correlated and thus neglected.

**Table 5**. Simulation results for the calculation of the $k_{\mathrm{Co,Cs}}$ factor

| S-Cs ($^{137}$Cs) | | | | | | | |
|---|---|---|---|---|---|---|---|
| $K_{\mathrm{air}}$ | | $D_{\mathrm{Cs}} \cdot m_{\mathrm{air}}$ | | $M_{\mathrm{Cs}}$ | | $N_{K,\mathrm{Cs}}$ | |
| Value / Gy cm$^2$ | *u* (%) | Value / MeV cm$^2$ | *u* (%) | Value / C cm$^2$ | *u* (%) | Value / Gy C$^{-1}$ | *u* (%) |
| 3.1177 x 10$^{-12}$ | 0.010 | 2.6448 × 10$^{-2}$ | 0.018 | 1.2473 × 10$^{-16}$ | 0.018 | 2.4996 × 10$^4$ | 0.021 |
| S-Co ($^{60}$Co) | | | | | | | |
| $K_{\mathrm{air}}$ | | $D_{\mathrm{Co}} \cdot m_{\mathrm{air}}$ | | $M_{\mathrm{Co}}$ | | $N_{K,\mathrm{Co}}$ | |
| Value / Gy cm$^2$ | *u* (%) | Value / MeV cm$^2$ | *u* (%) | Value / C cm$^2$ | *u* (%) | Value / Gy C$^{-1}$ | *u* (%) |
| 5.3552 × 10$^{-12}$ | 0.010 | 4.5973 × 10$^{-2}$ | 0.014 | 2.1681 × 10$^{-16}$ | 0.014 | 2.4700 × 10$^4$ | 0.017 |

Applying equation (2) to the last column, the correction factor is determined as: $k_{\mathrm{Co,Cs}} = 0.9882(3)$, where, the number in parentheses represents the standard uncertainty in the last significant digit due to statistical fluctuations alone.

Potential variations in the $k_{\mathrm{Co,Cs}}$ correction factor due to (i) axial beam non-uniformity, (ii) photon spectra of each beam quality (S-Cs and S-Co), (iii) photon interaction cross-sections, and (iv) geometric model of the chamber, were evaluated via sensitivity analyses as described in sections 2.2.1 and 2.2.2. These contributions are incorporated as additional uncertainty components, as summarized in table 6.

**Table 6**. Uncertainty components of the $k_{\mathrm{Co,Cs}}$ factor

| Evaluated uncertainty components | $u$ (%) (1) |
|---|---|
| Statistical uncertainty | 0.03 |
| $K_{\mathrm{air}}$ calculation method (2) | 0.12 |
| Cross-sections (3) | 0.03 |
| Photon spectra | 0.20 |
| Axial beam uniformity | 0.15 |
| PTW 32002 chamber model | 0.30 |
| Combined standard uncertainty | 0.41 |

*(1) Relative standard uncertainty*

*(2) Uncertainty calculated from the variation of $K_{air}$ values obtained using different methods. Uncertainties from the cross-sections themselves are not included, as they are strongly correlated in the ratio of equation (2).*

*(3) Uncorrelated fraction.*

Considering the combined uncertainty: $\boldsymbol{k_{\mathrm{Co,Cs}}}$**= 0.988 ± 0.008 (*k* = 2).**

To ensure adequate statistical uncertainty, $1 \times 10^8$ and $2 \times 10^8$ photon histories were simulated for air kerma and absorbed dose calculations, respectively.

### 3.2 Experimental validation of the calculated $k_{\mathrm{Co,Cs}}$ factor

#### 3.2.1 Validation against historical calibration data

Experimental verification of the correction factor $k_{\mathrm{Co,Cs}}$ was performed using certificates from successive chamber calibrations. The PTW 32002 (00345) chamber has been calibrated in-house against the PTW 32005 (00047) chamber each time the latter was calibrated at a Primary Standard Dosimetry Laboratory (PSDL), prior to the establishment of our own $^{137}$Cs primary standard. These chamber calibrations were consistently performed for both beam qualities.

The standard uncertainty due to the long-term stability of the PTW 32002 (00345) chamber between 2012 and 2020 is 0.2%. This relatively low uncertainty does not affect the correction factor, as the calibrations for both beam qualities were conducted within a few hours of each other.

Table 7 summarizes the calibration coefficients for the PTW 32002 (00345) chamber obtained over the years (the corresponding calibration certificate code is indicated in parentheses). The last row includes the $k_{\mathrm{Co,Cs}}$ factor values calculated for each calibration. Uncertainties refer to the expanded uncertainty ($k$ = 2). The uncertainty of the $k_{\mathrm{Co,Cs}}$ factors excludes those components that are fully correlated during the calibration process in both beam qualities. Specifically, the following correlated components were not considered for the ratio:

- At the PSDL level: the uncertainty of physical constants, Type B uncertainties of the correction factors for the primary standard, the electrometer calibration, Type B uncertainties associated with leakage current corrections, and the long-term stability of the transfer standard.
- During in-house calibrations: the long-term stability of the reference chamber (PTW 32005, 00047), as well as Type B components associated with the electrometer calibration and the determination of influence quantities (pressure, temperature, and humidity) during the internal calibration of the PTW 32002 (00345) chamber.

These factors cancel out when calculating the ratio of the calibration coefficients, meaning the reported uncertainty for $k_{\mathrm{Co,Cs}}$ primarily reflects the non-correlated effects and the short-term statistical fluctuations of the measurement system.

**Table 7**. Calibration coefficients of the PTW 32002 (00345) chamber for S-Cs and S-Co radiation qualities

| Year (certificate code) | Beam Quality | $N_K$ / Gy C$^{-1}$ | $U$ / Gy C$^{-1}$ | $k_{\mathrm{Co,Cs}}$ |
|---|---|---|---|---|
| 2012 | S-Cs | $2.548 \times 10^4$ | $4.077 \times 10^2$ | $0.987 \pm 0.013$ |
| (I123/LMRI/GP/605) | S-Co | $2.515 \times 10^4$ | $4.024 \times 10^2$ | |
| 2014 | S-Cs | $2.532 \times 10^4$ | $4.026 \times 10^2$ | $0.985 \pm 0.014$ |
| (I337/LMRI/GP/1262) | S-Co | $2.495 \times 10^4$ | $4.167 \times 10^2$ | |
| 2017 | S-Cs | $2.534 \times 10^4$ | $4.029 \times 10^2$ | $0.986 \pm 0.015$ |
| (I671/LMRI/GP/2691) | S-Co | $2.499 \times 10^4$ | $4.448 \times 10^2$ | |
| 2019 | S-Cs | $2.502 \times 10^4$ | $3.953 \times 10^2$ | $0.984 \pm 0.014$ |
| (I857/LMRI/GP/3379)(1) | S-Co | $2.462 \times 10^4$ | $4.136 \times 10^2$ | |

*(1) The 2019 coefficients reflect PSDL reference updates in accordance with the ICRU Report 90 [8].*

### 3.2.2 International validation: the EURAMET.RI(I)-S19 supplementary comparison

An additional independent validation was provided by the results of the EURAMET.RI(I)-S19 supplementary comparison, recently published in the BIPM Key Comparison Database (KCDB) [16]. In this exercise, CIEMAT participated as a primary laboratory, benchmarking its results against the NPL primary standard for $^{60}$Co beams at radiation protection levels. For this purpose, CIEMAT established its traceability through its own $^{137}$Cs primary standard [1, 2], applying the beam-quality correction factor $k_{\mathrm{Co,Cs}}$ derived in this work for the PTW 32002 (00345) transfer chamber.

The results for CIEMAT, analyzed in terms of the degrees of equivalence relative to the comparison reference values, demonstrate excellent agreement. The consistency of these international results conclusively validates the accuracy of the calculated $k_{\mathrm{Co,Cs}}$ factor.

## 4 Conclusions

An independent and robust air kerma traceability chain for $^{60}$Co radiation protection levels has been successfully established at the LMRI-CIEMAT. This methodology effectively overcomes the technical limitations associated with the low signal-to-noise ratio of small-volume primary standards in low dose-rate beams. By utilizing a high-volume secondary standard chamber (PTW 32002) calibrated against the $^{137}$Cs primary standard, the laboratory has eliminated its reliance on external primary calibrations for $^{60}$Co protection levels.

The beam-quality correction factor, $k_{\mathrm{Co,Cs}}$ , calculated via Monte Carlo simulations with the EGSnrc code, proved to be highly accurate. The cancellation of the air mass (chamber volume) in the calculation formalism reduces the requirements for the ionization chamber to be used as transfer standard; furthermore, the cancellation of the average energy required to produce an ion pair significantly reduces the final uncertainty budget of the traceability chain.

The reliability of this independent method was internationally validated through CIEMAT's successful participation in the EURAMET.RI(I)-S19 supplementary comparison, now officially registered in the BIPM KCDB. CIEMAT's participation as a primary laboratory confirms that the implemented traceability, based on the $^{137}$Cs primary standard and the calculated $k_{\mathrm{Co,Cs}}$ factor for a well-characterized ionization chamber, is equivalent to international benchmarks.

## Acknowledgments

The authors are grateful to Daniel Gijón Tiradas from the Energy Materials Division at CIEMAT for his technical support in acquiring the radiographic images of the transfer ionization chamber.